\documentclass[epsfig]{article}
\usepackage{graphicx}

\newcommand{\be}{\begin{equation}}
\newcommand{\ee}{\end{equation}}

\begin{document}

\title{\bf Spectral Libraries for Analyzing Spectra of Low-Metallicity Galaxies}
\author{I. Hubeny\footnote{University of Arizona, 
Tucson; AZ; ihubeny.astr{\tt @}gmail.com}, and
Sara R. Heap\footnote{NASA Goddard Space Flight Center, Greenbelt, MD, emerita}}
\maketitle

\begin{abstract}
We present a set of isochrone-tailored spectral libraries for analyzing composite spectra of low-metallicity massive star clusters or starburst galaxies.  Specifically, we have computed non-LTE spectra for stars of all initial masses for isochrones at metallicities, $Z=0.006,\, 0.002,\,{\rm and}\, 0.0004$, with and without considering rotation. These isochrones were constructed by the Geneva group (Ekstr\"{o}m et al., 2011; Georgy et al., 2013; Groh et al., 2019;  Eggenberger et al. 2020). We also present a Python program for integrating individual spectra for an arbitrary initial mass function.
\end{abstract}

\section{Introduction}

Predicting the composite spectrum of a stellar system, e.g. starburst cluster or galaxy composed of stars of essentially same age but a wide range in  initial masses, involves several basic ingredients:
\begin{itemize}
\item evolutionary models and consequent isochrones  for a set of ages of the system;
\item synthetic spectra of stars making up the isochrone; and
\item an initial-mass function, i.e. distribution of stellar initial mass.
\end{itemize}

Isochrones describe the dependence of basic stellar parameters on initial mass, age, and rotation. 
 These basic parameters include effective temperature, surface gravity, radius (or bolometric luminosity), and chemical abundances of most important chemical species.
Isochrones for stellar systems of various initial metallicities are available from the Geneva group (Ekstr\"{o}m et al., 2011; Georgy et al., 2013; Groh et al., 2019; Eggenberger et al. 2021). Spectra for individual stars are usually taken from general-use spectral libraries, 
which typically contain spectra for a discrete set of effective temperatures, $T_{\rm eff}$, surface gravities, $\log g$, 
and metallicities. By "metallicity," we adopt a single value representing the abundance ratio of oxygen to hydrogen compared to the solar abundance ratio.  The solar metallicity is taken as $Z_\odot = 0.014$. 

There are, however, several drawbacks to using such spectral libraries:
\begin{itemize}
\item One needs to perform a number of 2- or 3-dimensional interpolations to determine the stellar spectrum for stellar parameters (effective temperature, surface gravity, metallicity) stipulated by the isochrone table;
\item The parameter space covered by the library may be insufficient. For instance, some $(T_{\rm eff} - \log g)$  pairs required by the isochrone table are beyond the range of the adopted spectral library.

 \item The evolution of rotation-induced surface abundances affects some species differently than others. An obvious example is an increase of nitrogen and decrease of carbon abundance with age. Most pre-constructed spectral libraries do not typically consider selective abundance patterns.
\item For certain studies, existing spectral libraries may have  insufficient spectral resolution needed for intended study such as line-profile analysis.
\end{itemize}

These problems are readily lifted if a spectral library is tailored for a given isochrone, so there are no gaps in the parameters space, and no interpolations are needed. The spectral resolution may also be chosen to enable detailed and accurate analysis. Current computer memory, both internal, as well as external storage is readily available and is becoming increasingly cheaper, so this concern is no longer serious.

\section{New isochrone-tailored spectral libraries}

In view of the reasons outlined above, we present here a set of synthetic spectral libraries specifically tailored to Geneva isochrones, which are available from the university's SYCLIST website (Ekstr\"{o}m et al). 
We have constructed synthetic spectra for all initial masses, taking the basic stellar parameters
($T_{\rm eff}$, $\log g$, chemical abundances, and luminosity) from the isochrone tables. We leave a more detailed description of the modeling procedures (underlying model atmospheres; details of spectrum synthesis) to Appendix A.
Here, we only stress that the presented spectra are {\em photospheric} spectra, i.e. a possible stellar wind is not taken into account.
The relevant quantities extracted from the isochrone table are: effective temperature, surface gravity, luminosity, radius, and current chemical abundances of the most important species, namely He, C, N, O, Ne, and Al, and rotational velocity if rotation is being considered in the stellar evolution.

Our library currently includes spectra for three different initial metallicities, $Z=0.006$, $Z=0.002$, and $Z=0.0004$, each with and without considering rotation. In all cases, the only value of rotation considered is 
$V/V_{\rm crit} = 0.4$, i.e. $\Omega/\Omega_{\rm crit}=0.568$. In all cases, the isochrones include 5 ages, 
$\log ({\rm age}) =6.0, 6.3, 6.5, 6.7$, and $7.0$ years. Altogether, there are 30 sets of spectra stored in 30 directories. 
The naming convention of the isochrones is simple and self-explanatory. For instance,
{\tt Z0002t6.7R} contains a set of spectra for the metallicity, $Z=0.002$, at age $t = 10^{6.7}$ years, and with rotation (label {\tt R}). The isochrones are summarized in Table 1, which also shows the number of individual initial masses considered for a given isochrone.
\begin{table}
\caption{Number of considered initial masses in the  individual isochrone sets}
\begin{center} 
\begin{tabular}[b]{l r r r r r}
\hline
Isochrone Name/Age  & 6.0 & 6.3 & 6.5 & 6.7 & 7.0 \\
\hline
Iso\verb|_|Z006t[{\tt age}] & 51 & 51 & 115 & 98 & 88 \\
Iso\verb|_|Z006t[{\tt age}]R & 51 & 51 & 104 & 101 & 97 \\
Iso\verb|_|Z002t[{\tt age}] & 51 & 51 & 114 & 124 & 136 \\
Iso\verb|_|Z002t[{\tt age}]R & 51 & 51 & 51 & 119 & 126 \\
Iso\verb|_|Z0004t[{\tt age}] & 51 &  51 &124 & 141 & 161 \\
Iso\verb|_|Z0004t[{\tt age}]R & 51 & 51 &123 & 128 & 145 \\
\hline
\end{tabular}
\\[2.5pt]
{\small
Here, [{\tt age}] is a 3-elements string representing log (age) in years, e.g., {\tt 6.0}, {\tt 6.3}, etc.
}
\end{center}
\end{table}
An isochrone directory is composed of a set of synthetic spectra for all initial masses in the corresponding Geneva table, 
together with an overview table with the name {\tt{*.tab}}. In this example, {\tt Z002t6.7R.tab}, lists a subset of values from the corresponding Geneva isochrone table which are directly relevant to spectrum synthesis. Individual columns give:
\begin{itemize}
\item initial mass [in $M_\odot$]
\item effective temperature [K];
\item $\log g$ - gravity acceleration at the surface [cm s${}^{-2}$]
\item $\log L$ - logarithm of the total luminosity [in $L_\odot$]
\item $v_{\rm eq}$ - equatorial rotational velocity [km s${}^{-1}$]
\item stellar radius [in $R_\odot$]
\item logarithm of the mass loss rate [in $M_\odot {\rm yr}^{-1}$].
\item GA-Ed, Eddington Gamma
\item the last 5 columns are the abundances of He, C, N, O, and Ne relative to the solar abundance.
These abundances were derived from the mass fractions of elements given by the Geneva isochrone tables.
Our adopted values for the solar abundances, expressed as $\log_{10}(N_{\rm atom}/N_H) +12$, are:
10.93, 8.39, 7.78, 8.66, 7.84, for He, C, N, O, and Ne, respectively
\end{itemize}
Figure 1 shows a screenshot of the first several lines of one such file.
\begin{figure}[h]
\begin{center}
\label{fig1}
\includegraphics[width=5in]{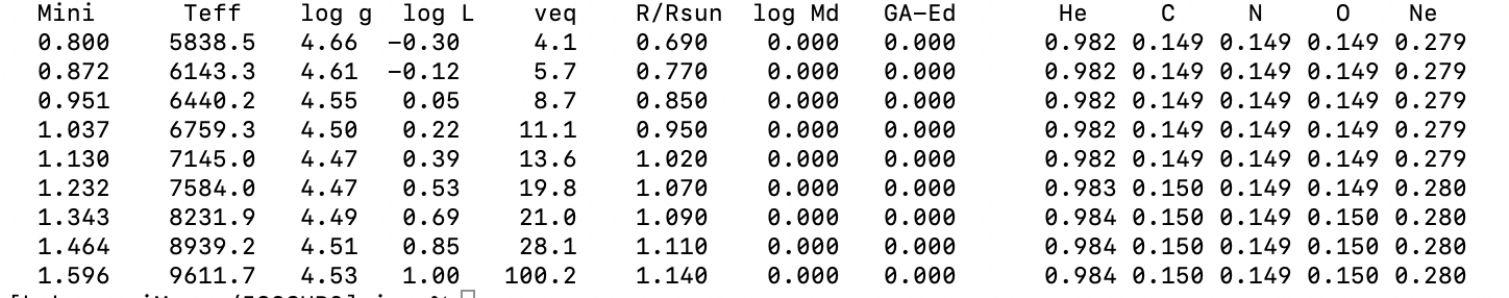}
\caption{A screenshot of the beginning of a sample {\tt *.tab} file.}
\end{center}
\vspace{-1em}
\end{figure}


All spectra have a wavelength spacing corresponding to a resolving power, $R=20,000$, throughout the full wavelength range of 200 to 10,000 \AA, which enables a large variety of studies.
The naming convention for individual models is given by an example. For instance, the file

 {\tt Z002t6.7Rm27.583.spec}

\noindent covers the synthetic spectrum  for a star with the initial mass $M_{\rm ini} = 27.583\,M_\odot$.
Each such file contains three or four columns, depending on whether the isochrone is constructed without considering rotation (3 columns), or with rotation (4 columns).

For the spectra without rotation, the individual columns are:
\begin{itemize}
\item wavelength [\AA];
\item monochromatic luminosity [erg s${}^{-1} {\rm \AA}^{-1}$]
\item monochromatic luminosity in the continuum (same units).
\end{itemize}
We chose to specify the continuum spectrum at all wavelengths. This is not strictly necessary because the continuum  is very smooth compared to the true synthetic spectrum, but this arrangement is very convenient because, again, it avoids interpolating in wavelength.

For the spectral set corresponding to the isochrone constructed with rotation. the meaning is analogous, but requiring
an additional explanations, namely
\begin{itemize}
\item wavelength [\AA];
\item monochromatic luminosity [erg s${}^{-1}{\rm \AA}^{-1}$], for a non-rotating star.  Although the isochrone assumes that stars rotate, this luminosity  is included in order to allow the user to consider
some other treatments of rotation, or to adopt some more sophisticated treatment of distribution of rotational velocities and rotational axes.
\item monochromatic luminosity in the continuum;
\item monochromatic luminosity, computed taking rotation into account, with the rotational velocity taken as $v \sin \approx 0.64\, v_{\rm eq}$, which represents a mean of projected rotational velocities assuming the rotation axes are distributed randomly (see also Appendix A).
\end{itemize}

As we will describe  in \S\,3, and in Appendix B, we provide not only the individual initial-mass spectra, but also the IMF-integrated spectra, called  {\tt imfinteg} spectra.
Some users may want first to examine the {\tt imfinteg} files which contain the co-added spectra from all initial masses of an isochrone with a Salpeter IMF applied. Altogether, there are 30 files for 3 metallicities (Z006, Z002 and Z0004), 5 ages (log age=6.0, 6.3, 6.5, 6.7, and 7.00), and 2 rotational status (non-rotating or rapidly rotating). Inter-comparison of the spectral isochrones of the five different ages enables one to “watch” how aging affects lines of interest. Comparison of spectral isochrones of rotating stars with that of non-rotating stars of the same age can show surprising differences, because rapidly rotating stars have longer main-sequence lifetimes.
 
The format of {\tt imfinteg} spectra has 4 columns for non-rotating stars giving: wavelength, total monochromatic luminosity, continuum luminosity , and monochromatic luminosity normalized to the continuum. {\tt Imfinteg} spectra for rapidly rotating stars has 5 columns, with the 5th (rightmost) column giving the total monochromatic luminosity without rotational broadening applied. It is useful in untangling differences due to composition from differences in line strength. The monochromatic luminosity is expressed in units of erg/s/$M_\odot/\AA$. 
 
There is much to be learned from {\tt imfinteg} spectra, but sooner or later, users will want to consult the individual initial-mass spectra from which the {\tt imfinteg} spectra are formed. These spectra are invaluable in learning why a spectral line is strong or weak, or why the profile of spectral line has the shape that it does.

\section{Creating Custom Integrated Spectra}

Individual spectra of stars with given initial masses are the basic ingredient of the present library. However, to obtain astronomically interesting results, we have to provide a mechanism to generate integrated spectra for the whole
stellar system. To this end, we present a Python program, {\tt iso.py} that contains the routine, {\tt integ}. This routine integrates the individual spectra of a given isochrone, for a specified initial mass function (IMF), represented by a 
power law, and for specified lower and upper initial mass cutoffs. Some details of the integration procedure are presented in Appendix B.

The program requires that all the {\tt *.spec} files and the corresponding {\tt *.tab} file are located in the same subdirectory, and all the filenames are the same as in the standard distribution of the files; otherwise the program would not work.

We stress that the resulting IMF-integrated spectrum is normalized to one solar mass.

\smallskip

The program is called with several parameters:
\begin{itemize}
\item[d0] - string, optional, if set,  specifies the directory where the spectral set is located. Default is {\tt './'}, i.e. the current directory.
\item[pref] - string that specifies the prefix of the file names with the spectra, i.e. the part of the filename common for all initial mass spectra. In other words, it is the part of the filenames before the mass indication, e.g.,  the string before {\tt m43.756.spec} 
\item[alpha] - float, optional, sets the IMF power law index. If not specified, the default  is {\tt alpha}=$-2.35$, i.,e. the Salpeter initial mass function (see Appendix B).
\item[param] - integer, optional, Sets the index of the column of the given spectral file that contains the integrated quantity. The default is {\tt param}=1, i.e. one integrates the full synthetic spectrum. For {\tt param}=2, the program  integrates the continuum luminosity.
\item[mmin] - float, optional. If set, it specifies the minimum mass for integrations (in $M_\odot$). The default is   the lowest mass included in the isochrone table (typically 0.8).
\item[mmax] - float, optional. Analogous to {\tt mmin}, but  for the maximum mass. The default is {\tt mmax} given by the maximum initial mass of the isochrone table.
\item[wmin] - float, optional. Sets the lowest wavelengths for the integrated spectrum.
       Default is wmin=0
\item[wmax] - float optional. Analogous for the highest wavelength. Default 10001.
\end{itemize}

The program returns three arrays:\\[2pt]
\noindent {\tt wave} - wavelengths [in \AA]\\[1pt]
\noindent {\tt spec}  - integrated spectrum (normalized to 1 solar mass) for these wavelengths 
[erg s${}^{-1}$ \AA${^{-1}M_\odot^{-1}}$]\\[1pt]
\noindent {\tt cont} - Normalized luminosity, i.e the integrated luminosity divided by integrated luminosity in the
continuum.\\[2pt]
The integrated spectra are normalized to 1 solar mass, so the total luminosity of a stellar system is obtained
by multiplying  {\tt spec} or {\tt cont} by the total mass, in units of the solar mass.
The program also produces a file with the name composed of {\tt pref} and {\tt .tot} that contains the integrated spectrum.\\
The file has four or five columns, depending one whether the isochrone is constructed for non-rotating or rotating stars,
namely\\[2pt]
\noindent $\bullet$ wavelength [\AA]\\
\noindent $\bullet$ integrated monochromatic luminosity [erg s${}^{-1}$\AA${}^{-1}M_\odot^{-1}$]. For isochrones with rotation, it is an integral of {\em rotated} spectra;.\\[2pt]
\noindent $\bullet$ integrated monochromatic  luminosity  in the continuum (same units);\\[2pt]
\noindent $\bullet$ integrated monochromatic  luminosity in the continuum;\\[2pt]
The files for isochrones  with rotation contain in addition the fifth column,\\[3pt]
\noindent $\bullet$ the integrated monochromatic luminosity evaluated without taking into account stellar rotation..

\medskip

For example, a command (in plain Python)%
\noindent
\begin{verbatim}
>>> import iso
>>> wave,spec,cont = iso.integ(pref='Z002t6.0')
\end{verbatim}
produces an integrated spectrum for the {\tt Z00t6.0} isochrone for the Salpeter IMF, $\alpha=-2.35$, and for all initial masses contained in the table. Besides the returned parameters {\tt wave, spec, cont}, the program also produced file 
{\tt Z002t6.0.imfsalp}, whose content is described above.

This example assumes that 
one runs Python in the directory that includes the spectral files for the appropriate isochrone.
The spectra in this
directory have filenames, for instance,  {\tt Z002t6.0m0.8.spec}, {\tt Z002t6.0Rm0.884.spec}, etc. One may then simply plot the integrated spectrum as\\[4pt]
\noindent{\tt >>> import matplotlib.pyplot as plt}\\
\noindent{\tt >>> plt.plot(wave,spec)}\\[2pt]
\noindent or with a number of appropriate keyword parameters.

For isochrones constructed with rotation, we stress that the non-rotated spectrum is given by the 2nd column, while the rotated spectra by the fourth column of the individual {\tt *.spec} files. Therefore, the analogous integrated spectrum that takes into account rotation, and computed only between 1100 and 1300 \AA, is produced by\\[3pt]
{\tt >>> iso.integ(pref='Z002t6.0',param=3,wmin=1100,wmax=1300).}\\[2pt]

To integrate over the masses between 1 and 30 $M_\odot$, and with the exponent $\alpha=-1.6$, and
issues a command\\[4pt]
\noindent{\tt >>> w,s,c = iso.integ(pref='Z002t6.0R',alpha=-1.6,mmin=1,mmax=30)}\\[4pt]
\noindent and the relative spectrum can be plotted as\\[4pt]
\noindent{\tt >>> plt.plot(w, s/c)}

The program  {\tt isp.py} also contains a simple routine {\tt specplot} which plots several spectra contained in the package (or elsewhere). It is called with two parameters:
\begin{itemize}
\item[files]  - string that specifies a list of lines to be plotted in the syntax of Linux command {\tt ls}. For instance, for
{\tt files = 'Z002t6.7Rm3[0-9]*.spec'} one plots the content of all files that are selected by means of the command
{\tt ls Z002t6.7Rm3[0-9]*.spec}; that is spectra for all initial masses between 30 and 40 $M_\odot$ of the isochrone {\tt Z002t6.7R}.
Note: if the program is not called from the directory where the files are located, one has to specify the whole path.
\item[param] -- integer, optional. Sets the index of the parameter to be plotted. Default {\tt param=1}, i.e. the full spectrum.
With {\tt param=2} one plots the relative spectrum.
\end{itemize}

\section{Obtaining the spectral libraries and codes}

The files can be downloaded from:\\[-4pt]

   {\tt https://www.as.arizona.edu/}$\,\widetilde{\,\,}\,${\tt hubeny/isochrones}\\

At this site, there are several types of files/directories:\\[2pt]:

-- the TAB directory, which contains a table of parameters of each isochrone in the library as taken or derived from the Geneva isochrones. The size of a tab file ranges from 5.2K to 16K.

-- 30 gzipped tar directories for the 30 isochrones, labeled by the name of the isochrone.   Each contains the full spectrum of each mass-point of the isochrone, e.g. 51 spectra comprising the Z004t6.0. The file names of individual spectra give the stellar mass and have the extension, .spec, e.g. Z004t6.0m3.658.spec.

--  the same 30 isochrone directories, with the name giving the isochrone and stellar mass, so that a user can access individual  {\tt *.spec} files, rather than downloading the whole, large set of spectra for the given age and metallicity;

-- the IMFSALP directory containing the gzipped IMF-integrated spectrum for all 30 isochrones, with the Salpeter power-low exponent {\tt 	alpha=-2.35}. The individual files have names of the isochrone with the extension, {\tt *.imfsalp}, e.g. {\tt Z002t6.7R.imfsalp}. This directory is simply an example of what can be obtained with the accompanying Python program {\tt iso.integ}.

-- the IMF1.35 directory, which contains the gzipped IMF-integrated spectrum of each of the 30 isochrones assuming a top-heavy IMF with {\tt alpha=-1.35}.

- Python program, {\tt iso.py}, which contains the routine, {\tt integ}, plus some utility routines;

-- Manual, file duplicating this document, {\tt manual.pdf}.

\section*{Appendix A: Model atmospheres and synthetic spectra}

We make use of NLTE metal line-blanketed model atmosphere grids OSTAR2003 (Lanz \& Hubeny 2003), and BSTAR2007 (Lanz \& Hubeny 2007) that contain a large number of models for various metallicities. The BSTAR grid covers effective temperatures between 15,000 and 30,000 K, and the OSTAR grid between 27,500 and 55,000 K.  The highest $\log g$ is 4.75, while the lowest $\log g$ depends on effective temperature, and is essentially determined as the lowest gravity for which the model is stable (close to the Eddington limit).

Since some basic stellar parameters, $T_{\rm eff}$ and $\log g$, stipulated in the isochrone tables are outside the range covered by the OSTAR and BSTAR grids, we have computed such models using the same procedure as used for original OSTAR and BSTAR grids.  We have also extended the set of NLTE metal line-blanketed model to lower effective temperatures. The new model grid is called LBASTAR (for late-B and A stars), and will be described in a future paper (Hubeny et al., in prep.).
The new models are computed using the program, {\sc tlusty}, (Hubeny \& Lanz 1995), in its newest version 
{Hubeny \& Lanz 2017; Hubeny et al. 2021).

Individual spectra are constructed as follows: First, we find the appropriate metallicity set from the OSTAR, BSTAR, and LBASTAR grids. In our case, these are T-models ($Z/Z_\odot = 0.1$ and L-models ($Z/Z_\odot = 0.5$.  We find  values of $T_{\rm eff}^{\rm grid}$ 
and $\log g^{\rm grid}$ that represent the closest lower and higher values for the given $T_{\rm eff}$ and $\log g$, 
and interpolate the four grid model structures to a new one at the $T_{\rm eff}$ and $\log g$ given by the isochrone table, and  compute the synthetic spectrum for this model.
As shown by Lanz \& Hubeny (2003), the most accurate way to obtain synthetic spectra for a model atmosphere with 
parameters between the grid values of $T_{\rm eff}$ and $\log g$ considered in the grid is first to interpolate to the 
atmospheric structure (temperature, density, atomic level populations) to get a new model atmosphere, and with this interpolated model atmosphere to calculate a synthetic spectrum, possibly with modified abundances of chemical elements, provided that the modifications of abundances are relatively small (say, up to about 0.3 dex; which is the case here).
We use computer program {\sc synspec} (Hubeny \& Lanz 2011, 2017).

For basic parameters of stars considered in the isochrones with rotation, the synthetic spectra are 
constructed as follows. First, the synthetic spectrum is computed for a unit area (1cm${}^2$) of the stellar surface, 
exactly as for non-rotating stars.  The isochrone tables give for each initial mass the equatorial rotation 
velocity, the polar $\log g$, and oblateness. The evolutionary models thus take into account a distortion 
of the stelar shape due to rotation.
However, as the isochrone tables show, these effects are rather small, and so we neglect the non-sphericity 
and possible dependence of $T_{\rm eff}$ and $\log g$ on the latitude on the stellar surface.  The rotational 
convolution is thus performed in the standard way (e.g. Gray, 1976, Chap. 11). Next, we assume that the stellar 
rotation axes are distributed randomly, so that for the rotational convolution we use an average value of 
$\sin i$ between 0 and $\pi$, namely $v_{\rm eq}\, 2/\pi \approx 0.64\, v_{\rm eq}$, where $v_{\rm eq}$ is 
the equatorial velocity given in the original isochrone tables.

\section*{Appendix B: IMF-Integrated spectra}

The synthetic spectrum of the whole stellar system is obtained by a standard procedure that we describe very briefly here. It is based on adopting  an initial mass function (IMF). For simplicity, we illustrate the procedure for the Salpeter (1955)  initial mass function where the IMF is defined by
\be
N(m)\, dm = w_0\, m^\alpha\, dm,
\ee
where $N(m) dm$ is the number of stars with masses in the range $(m, m+dm)$, and
$w_0$ is a normalization constant;
 $\alpha$ is the power-law exponent; for the Salpeter IMF, $\alpha=-2.35$ for $m > 0.5\, M_\odot$. 
 Another expression of Eq. (1) is that $dN/dm \propto m^\alpha$.
 Note that the usually quoted value of the
 Salpeter power-law exponent is $\alpha=-1.35$ which refers to the total {\it mass}  of stars in the elementary range, $m N(m)$.
 
 To integrate the set of individual spectra 
 to obtain the spectrum of the whole system as
 \be
 F_\lambda^{\rm tot} = \int_{m_{\rm min}}^{m_{\rm max}} N(m) F_\lambda(m) dm \approx
 w_0 \sum_i F_\lambda(m_i)\, w(m_i) N(m_i),
\ee
where the second, approximate, equality refers to a discretization of the integral over $m$, where $w_i$ is the appropriate 
integration weight. The normalization constant is determined by the condition on the total mass of the system,
\be
M = \int_{m_{\rm min}}^{m_{\rm max}} N(m) m\, dm = 
w_0  \int_{m_{\rm min}}^{m_{\rm max}} m^{1+\alpha} dm =
w_0 \left( m_{\rm max}^{2+\alpha} - m_{\rm min}^{2+\alpha} \right) / (2+\alpha),
\ee
and therefore 
\be
w_0 = (2+\alpha)/ \left( m_{\rm max}^{2+\alpha} - m_{\rm min}^{2+\alpha} \right) M \equiv w_{00} M.
\ee
The normalization constant $w_{00}$ has thus the meaning thats its application produces an IMF-integrated spectrum normalized to 1 solar mass.

This procedure is used in the accompanied Python program  {\tt iso.integ}, described in \S\,3..

\section*{References}

\def\reference{\par \leftskip20pt \parindent-20pt\parskip4pt}
\noindent

\reference Eggenberger, P., Ekstr\"{o}m, S., Georgy, C., 2021, A\&A, 652, 137.

\reference Ekstr\"{o}m et al., https://www.unige.ch/sciences/astro/evolution/en/database/syclist/

\reference Ekstr\"{o}m, S., Georgy, C., Eggenberger, P., et al., 2011, A\&A, 537, 146.

\reference Georgy, C., Ekstr\"{o}m, S., Eggenberger, P., et al., 2013, A\&A, 558, 103.

\reference Gray, D. F., 1976, The Observation and Analysis of Stellar Photospheres, Wiley, New York.

\reference Groh, J. H.; Ekstr\"{o}m, S.; Georgy, C., et al.,2019, A\&A 627, 24

\reference Hubeny, I., \& Lanz, T.  1995, ApJ, 439, 875.

\reference Hubeny, I., \& Lanz, T. 2011, {\sc synspec}: General Spectrum Synthesis Program, Astrophys. Source Code Lib. ascl:1109.022.

\reference Hubeny, I., \& Lanz, T. 2017, A Brief Introductory Guide to
TLUSTY and SYNSPEC, arXiv:1706.01859.

\reference  Hubeny, I., Allende Prieto, C., Osorio, Y., \& Lanz, T., 2021,  {\sc tlusty} and {\sc synspec} User's Guide IV:
Upgraded Versions 208 and 54., arXiv:2104,02829.

\reference Lanz, T., \& Hubeny, I., 2003, ApJS, 146, 417.

\reference Lanz, T., \& Hubeny, I., 2007, ApJS, 169, 83.

\reference Salpeter, E., 1955, ApJ, 121, 161.

\end{document}